\documentclass[11pt]{article}
\usepackage[final]{acl}

\usepackage{times}
\usepackage{latexsym}
\usepackage{amsmath}
\usepackage{algorithm}
\usepackage{algorithmic}
\usepackage{booktabs}
\usepackage{subcaption}
\usepackage{graphicx}
\usepackage{array}
\usepackage{placeins}
\usepackage{xcolor}
\definecolor{figRed}{RGB}{204, 0, 0} 
\usepackage{float}
\usepackage{makecell}
\usepackage[T1]{fontenc}

\usepackage[utf8]{inputenc}

\usepackage{microtype}

\usepackage{inconsolata}

\title{Prompt-Induced Over-Generation as Denial-of-Service: A Black-Box Attack-Side Benchmark}

\author{
  Manu \quad
  Yi Guo \\
  Western Sydney University \\
  Sydney, Australia
  \And
  \shortstack[c]{Kanchana\\Thilakarathna} \quad
  \shortstack[c]{Nirhoshan\\Sivaroopan} \\
  The University of Sydney \\
  Sydney, Australia
  \AND
  Jo Plested \quad Tim Lynar \\
  University of New South Wales \\
  Canberra, Australia
  \And
  Jack Yang \quad Wangli Yang \\
  University of Wollongong \\
  Wollongong, Australia
}

\begin{document}
\maketitle
\begin{abstract}
Large Language Models (LLMs) can be driven into over-generation, emitting thousands of tokens before producing an end-of-sequence (EOS) token. This degrades answer quality, inflates latency and cost, and can be weaponized as a denial-of-service (DoS) attack. Recent work has begun to study DoS-style prompt attacks, but typically focuses on a single attack algorithm or assumes white-box access, without an attack-side benchmark that compares prompt-based attackers in a  black-box, query-only regime with a known tokenizer. We introduce such a benchmark and study two prompt-only attackers. The first is an Evolutionary Over-Generation Prompt Search (EOGen) that searches the token space for prefixes that suppress EOS and induce long continuations. The second is a goal-conditioned reinforcement learning attacker (RL-GOAL) that trains a network to generate prefixes conditioned on a target length. To characterize behavior, we introduce Over-Generation Factor (OGF): the ratio of produced tokens to a model’s context window, along with stall and latency summaries. EOGen discovers short-prefix attacks that raise Phi-3 to $\mathrm{OGF}=1.39\pm1.14$ (Success@$\ge2$: 25.2\%); RL-GOAL nearly doubles severity to $\mathrm{OGF}=2.70\pm1.43$ (Success@$\ge2$: 64.3\%) and drives budget-hit non-termination in 46\% of trials.

\end{abstract}

\section{Introduction}
Large Language Models (LLMs) can be forced into severe over-generation by delaying or preventing emission of the end-of-sequence (EOS) token \citep{gao2024denial, dong2024engorgio}. This is known as a denial of service (DoS) or ``sponge'' attack, where the aim is to overwhelm or exhaust LLM service resources \citep{gao2024denial, geiping2024coercing, zhang-etal-2025-crabs}. In DoS attacks, LLM output is degraded and computational costs increase \citep{zhang-etal-2025-crabs}. This increases inference latency and can make LLMs inaccessible to other users or tasks, and in extreme cases can lead to service disruption \citep{dong2024engorgio}. 

We focus on prompt-induced attacks in which a model generates excessive output in response to a crafted prefix that interferes with termination behavior. For example, \citet{geiping2024coercing} suppress an EOS token and drastically increase system response length via Greedy Coordinate Gradient (GCG) or string repetition. Further, \citet{dong2024engorgio} demonstrate that adversarial Engorgio prompts can cause LLMs to generate abnormally long outputs and increase computation costs. Note also that DoS attacks can deplete system resources without affecting output accuracy \citep{hasan2025spongeattackssensingai}. As such, detection and prevention of such attacks are key. 

To address the lack of standardized \emph{attack-side} evaluation for prompt-induced over-generation, we introduce a black-box, query-only (tokenizer-known) benchmark that compares prompt-based attackers under a consistent interface and decoding regime. While prior work has studied other forms of programmatic misuse that increase inference cost (e.g., over-reasoning slowdowns or interruption-style behaviors) \citep{kumar2025overthink, cui2025practical}, our benchmark focuses specifically on \emph{stopping-time} vulnerability: how easily an attacker can delay termination and inflate output length. We report consistent stopping-time metrics and attacker setups so results are reproducible and comparable across models.

Within this benchmark, we study two automated, prompt-only attackers for short- to medium-length prefixes. \textbf{EOGen} performs gradient-free evolutionary search over a restricted prompt space, serving as a strong black-box, query-only baseline. \textbf{RL-GOAL} is a goal-conditioned reinforcement learning attacker that learns to generate prefixes targeting specific continuation lengths. We evaluate attacks using \textbf{Over-Generation Factor (OGF)} and complementary stopping-time summaries including \textbf{stall rate}, enabling model-agnostic comparison across model families, sizes, and decoding settings while remaining strictly attack-side.

\section{Related Work}

\paragraph{Adversarial prompting and limits of alignment.}
A large body of work shows that LLM behavior can be steered into undesirable regimes via adversarial prompts, including universal or transferable suffixes found by automated search \citep{zou2023universal, liao2024amplegcg}. From a theoretical perspective, \citet{wolf2023fundamental} argue that for any behavior that occurs with non-zero probability, there exist prompts that can increase its likelihood, highlighting fundamental limitations of alignment under adversarial querying. Complementary empirical evidence suggests that safety alignment can be shallow and brittle, motivating evaluation settings where attackers can probe termination and stopping behavior rather than relying on single-case demonstrations \citep{qi2024safety}.

\paragraph{Automated prompt optimization for adversarial prompt construction.}
Our attackers are automated and prompt-only, which connects to prior work on discrete prompt optimization and trigger search. Early work on universal adversarial triggers and gradient-guided token search demonstrated that short token sequences can reliably elicit targeted behaviors and expose model vulnerabilities \citep{wallace-etal-2019-universal, shin-etal-2020-autoprompt}. More recent work scales these ideas to aligned LLMs using automated suffix construction and search procedures that can transfer across prompts and even to black-box model interfaces \citep{zou2023universal}. Closest in spirit to our \textbf{EOGen} baseline, \citet{guo-etal-2025-evoprompt} introduce an evolutionary framework for discrete prompt optimization that is gradient-free and operates through iterative population-based exploration. Similarly, \citet{deng-etal-2022-rlprompt} propose reinforcement learning for discrete prompt optimization, learning a policy that generates prompts by maximizing a task-defined reward, which conceptually motivates our \textbf{RL-GOAL} attacker as a goal-conditioned prefix generator.

\paragraph{Over-generation and LLM denial-of-service.}
Recent work has begun to study DoS-style attacks that exhaust inference-time resources by inducing excessive generation or delaying termination. Token-level prompt optimization can suppress EOS and drive extreme length inflation (e.g., GCG-style methods) \citep{tang2025dsgcg, liao2024amplegcg}, while crafted prompts such as Engorgio demonstrate prompt-only over-generation and increased compute cost \citep{dong2024engorgio}. Beyond prompt-only settings, training-time manipulations such as P-DoS can also alter termination tendencies and prolong generation \citep{gao2024denial}. In the broader security literature, ``sponge'' examples show that carefully constructed inputs can maximize latency/energy and induce worst-case inference behavior, including for language models \citep{shumailov-etal-2021-sponge}. Most closely related to our setting, \citet{zhang-etal-2025-crabs} introduce a black-box automated DoS prompt construction approach (AutoDoS) and report large latency amplification across models. Together, these findings motivate standardized, attack-side benchmarks that compare prompt-based attackers under consistent black-box (query-only) interfaces and stopping-time metrics.

\section{Method: Evolutionary Over-Generation Prompt Search (EOGen)}
\label{sec:method-eogen}

\subsection{Black-Box, Query-Only Over-Generation Setup}
\label{subsec:blackbox-setup}

We study standard next-token prediction with decoder-only LLMs in a text-completion setting. A prompt \(p\) is mapped to a token sequence \(\mathbf{x} = (x_1,\dots,x_m)\), and the model then autoregressively samples a continuation \(\mathbf{y} = (y_1,\dots,y_L)\) conditioned on \(\mathbf{x}\) and the previously generated tokens. The decoding process is truncated either when a special end-of-sequence symbol \(\langle\mathrm{eos}\rangle\) appears or when a pre-defined upper bound on the number of generated tokens is reached. We refer to \(L = |\mathbf{y}|\) as the continuation length. In our experiments (Appendix~\ref{subsec:decoding}), we fix this upper bound and the decoding hyperparameters for each victim model so that differences in behavior arise only from the attacker’s choice of prompt, up to stochasticity introduced by sampling.

Our threat model is a black-box, query-only, inference-time attacker. The adversary may (i) supply short textual prefixes \(p\) to the victim model and (ii) observe the resulting continuations. We therefore treat the victim as a query-only text-completion oracle: the evolutionary search never accesses gradients, hidden states, or logits, and all decisions are based solely on the output tokens (with an explicit penalty for prompt length). This contrasts with training-time poisoning attacks such as P-DoS, which assume control over the fine-tuning data, and with white-box inference-cost attacks that optimize prompts using gradient information \citep{gao2024denial}.

EOGen searches for short adversarial prefixes that induce extreme continuation lengths. We formulate prompt discovery as an evolutionary optimization over a discrete token space under a strict evaluation budget: each candidate prompt is scored by a length-based fitness function that rewards longer continuations while penalizing prompt length and EOS-related termination. Section~\ref{sec:metrics} then defines the diagnostic metrics we report to characterize attack severity (e.g., Over-Generation Factor, stall events, and latency), which we log for analysis alongside the fitness values used for search.

\subsection{Prompt Representation and Search Space}
\label{subsec:prompt-space}

EOGen operates directly in the discrete token space of the victim model. Let \(\mathcal{V}\) be the tokenizer vocabulary and let \(\mathrm{dec} : \mathcal{V} \rightarrow \Sigma^{*}\) denote the decoding function that maps token IDs to decoded strings over the byte alphabet \(\Sigma\). We construct a restricted subset \(\mathcal{S} \subset \mathcal{V}\) of word-like token IDs. Concretely, we decode each \(v \in \mathcal{V}\) and retain it if its decoded form passes a simple filter (ASCII, alphabetic characters only, and present in an English word list; we implement this using the \texttt{words} corpus in NLTK). This filter removes non-alphabetic tokens and many tokens that are not standalone English words, and biases the search towards prompts that resemble benign English words rather than obviously synthetic token sequences.

A candidate prompt is represented as a short sequence of token IDs \(\mathbf{t} = (t_1,\dots,t_\ell)\) with \(t_i \in \mathcal{S}\) and length \(\ell\) constrained to a small interval \([\ell_{\min}, \ell_{\max}]\). We use the same \(\ell_{\min}\) and \(\ell_{\max}\) for all evolutionary runs and instantiate them as 3 and 7 tokens respectively in our experiments (Appendix~\ref{subsec:eogen-hparams}). These constraints make gradient-free search tractable under a fixed query budget and keep candidate prefixes interpretable.
The resulting search space
\begin{equation}
\mathcal{T} = \bigcup_{\ell=\ell_{\min}}^{\ell_{\max}} \mathcal{S}^{\ell}
\end{equation}
therefore contains only short prompts built from English-like tokens. Each \(\mathbf{t} \in \mathcal{T}\) is passed directly to the victim model as the full input ID sequence, i.e., we set \(\mathbf{x}=\mathbf{t}\). The attacker is thus restricted to manipulating a small number of word-like tokens, while the remainder of the generation behavior is determined by the victim model and the fixed decoding regime described in Appendix~\ref{subsec:decoding}.

\subsection{Fitness Function}
\label{subsec:fitness-metrics}

Given a candidate prompt \(\mathbf{t} = (t_1,\dots,t_\ell)\) and its generated continuation \(\mathbf{y}\) of length \(L\), EOGen uses a scalar fitness function to guide the evolutionary search. The reward is designed to favor prompts that induce very long continuations while discouraging unnecessarily long prompts and early termination. Concretely, we define
\begin{equation}
R(\mathbf{t}) \;=\; L \;-\; \alpha \,\ell \;-\; \mathbf{1}_{\mathrm{eos}} \,\beta \,\bigl(B - i_{\mathrm{eos}}\bigr),
\end{equation}
where \(\alpha > 0\) controls a linear penalty on prompt length, \(B\) is the maximum generation budget (in tokens), \(\mathbf{1}_{\mathrm{eos}}\) is an indicator that the continuation contains an end-of-sequence token, and \(i_{\mathrm{eos}}\) denotes the zero‑based index of the first end‑of‑sequence token within the continuation \(\mathbf{y}\). In our implementation we set \(\alpha = 2\) and \(\beta = 5\), and set $B$ to the maximum allowed number of newly generated tokens per query (the decoding budget). We include a linear penalty on prompt length (scaled by $\alpha$) to discourage degenerate solutions that rely on excessively long prefixes, which can consume context and reduce transferability. We also penalize early EOS emissions (scaled by $\beta$) because suppressing EOS is a prerequisite for inducing over-generation. 
We choose $\alpha$ and $\beta$ to balance these objectives while preserving search diversity under a fixed query budget. Prompts that induce long continuations with no EOS (or an EOS very close to the budget) therefore receive high reward, while prompts that stop early or rely on long prefixes are strongly down-weighted.

\subsection{EOGen Algorithm}
\label{subsec:eogen-algorithm}

EOGen instantiates a simple evolutionary loop over the prompt space $\mathcal{T}$, similar in spirit to evolutionary prompt optimization methods \citep{guo-etal-2025-evoprompt}. At generation \(g\) we maintain a population $P_g$ of $P$ candidate prompts at generation $g$. 
At the start of a fresh run (i.e., when no checkpoint is loaded), we initialise $P_0$ around a short seed string (``In''): we tokenize the seed, discard any token IDs not in the allowed set $\mathcal{S}$, and then pad or truncate with uniformly sampled tokens from $\mathcal{S}$ so that each prompt length satisfies $\ell \in [\ell_{\min}, \ell_{\max}]$. 
This produces an initial population of short, English-like prompts before optimisation. 
If a checkpoint is available, we instead resume the population from the saved state.

At each generation we evaluate all prompts in \(P_g\). Given a prompt \(\mathbf{t}\), we call the victim model’s \texttt{generate} API under the fixed decoding configuration (Appendix~\ref{subsec:decoding}), obtain the continuation \(\mathbf{y}\), and compute the reward \(R(\mathbf{t})\) defined in Section~\ref{subsec:fitness-metrics}. We maintain the best prompt \(\mathbf{t}^\star\) and reward \(R^\star\) seen so far across generations.

Selection and variation follow a standard truncation-selection scheme. After evaluation, we sort \(P_g\) by reward and retain the top \(\rho P\) prompts as survivors. In each generation we repeatedly sample pairs of survivors and perform a one‑point crossover by choosing a random split point in the first parent and exchanging the remaining segment with the second parent. We then mutate the resulting prompt by independently replacing each token with probability $\mu$ with a randomly chosen word‑token from \(\mathcal{S}\). Finally, we enforce the length constraint \([\ell_{\min}, \ell_{\max}]\) by padding with random tokens or truncating as needed until a new population \(P_{g+1}\) of size \(P\) is obtained. Algorithm~\ref{alg:eogen} summarizes the overall procedure.

\section{RL-GOAL: Goal-Conditioned Reinforcement Learning Attacker}
\label{sec:rl-goal}
\subsection{Black-Box RL Setup}
\label{subsec:rl-blackbox}

RL-GOAL's \emph{RL interaction loop} operates in a query-only, inference-time setting analogous to EOGen. During training and evaluation, the environment interacts with the victim model only by invoking HuggingFace \texttt{generate} to obtain a fresh victim continuation and compute a \emph{terminal} reward (detailed in subsection~\ref{subsec:rl-reward}); we do not access victim gradients or hidden states in this RL loop.

We cast prompt construction as an episodic Markov decision process (MDP) \citep{sutton2018reinforcement} in which the agent incrementally builds a short prefix. At step $t$, the environment observation is a fixed-length view of the current prompt: the last 1024 token IDs of the concatenated seed prefix and agent-selected tokens, left-padded with zeros if shorter. Observations are a fixed-length 1024-token window; padding is used for shorter prompts and ignored when forming the policy context; actions are sampled with EOS masked and an ASCII ‘English-ish’ vocabulary mask. The seed prefix defaults to \texttt{"The following text is generated:"} but can be overridden at environment reset. During training, we optionally initialize the seed prefix from a prompt pool with a decaying mixture probability (Appendix~\ref{app:prompt-init}). The policy selects the next token from a restricted \emph{English-ish} subset of the vocabulary (a precomputed ASCII-based mask). To avoid trivial early stopping, the EOS action is masked during training rollouts; episodes truncate at a fixed horizon ($T{=}64$), with overshoot safeguards detailed in Appendix~\ref{subsec:rl-impl}. To keep training cost predictable and stabilize optimization under a fixed query budget, we bound the rollout horizon ($T{=}64$). We further include two overshoot guards trainer side early stopping at $1.2\times g_{\text{agent}}$ and an environment-side hard cutoff at $1.3\times g_{\text{agent}}$ which prevent runaway prompt construction while still allowing limited overshoot beyond the sampled goal.

At the start of each batch we sample a target continuation length $g$ from a curriculum whose support expands over training. Concretely, we sample $g \in [g_{\min}, g_{\max}(k)]$ with $g_{\min}=2048$ and an upper bound $g_{\max}(k)$ that increases with training while respecting (i) a global generation cap and (ii) a bootstrapping cap that limits the victim's generation budget early in training detailed in Appendix~\ref{app:rl-curriculum}. This curriculum improves optimization stability and sample efficiency by first training the policy on attainable targets, then gradually exposing it to long-context goals as training progresses. In parallel, the bootstrapping cap throttles the victim-side decoding budget early in training, avoiding wasted queries on long goals before they become feasible, and is relaxed in stages toward the full long-context regime (details in Appendix~\ref{app:rl-curriculum}).

To emphasize long-target behavior, we bias sampling toward the upper half of the interval. After the prefix is constructed, we query the victim once using a fixed decoding configuration presented in Appendix~\ref{app:rl-victim-query} and a per-episode token budget $B$, yielding a continuation length $L\le B$ from which the terminal reward is computed. In our implementation, $g_{\max}(k)=\min(G_{\max},c_k,b_k)$ and $B=\min(g,G_{\max},b_k)$, and we increase the upper-half sampling probability once $b_k$ reaches the long-context regime.

\subsection{Policy Architecture and Action Selection}
\label{subsec:rl-arch}

The RL-GOAL attacker is parameterized by a policy network \(\pi_\phi(a_t \mid s_t, g)\), which is the actor that produces the next-token action distribution \citep{sutton2018reinforcement, schulman-etal-2017-ppo} and a matching value network \(V_\psi(s_t, g)\) substantially reduces gradient variance and stabilizes learning \citep{schulman-etal-2017-ppo, schulman2016gae}. Both networks operate directly on token IDs via learned token embeddings. We condition on the target length by discretising \(g\) into a set of discrete length buckets. In our implementation, we partition the goal range (up to 16k tokens in our experiments) into buckets of width 256 tokens each (yielding 65 bucket indices from 0 to 64) and map each bucket index to a learned goal embedding. We use 256-token buckets as a pragmatic trade-off between conditioning resolution and parameter/coverage efficiency. With a maximum goal of 16k, this yields 65 discrete goal embeddings, keeping the goal-conditioning table small and ensuring each bucket is visited frequently during training. 
Finer discretization would increase the embedding dimensionality and reduce per-bucket data, while coarser discretization would blur distinct length targets.
This goal embedding vector is inserted at a dedicated position (prepended to the token sequence) and is used in both the policy and value networks. Concretely, after the transformer stack we read out the hidden state at the prepended goal position and project it to (i) vocabulary logits for $\pi_\phi$ and (ii) a scalar for $V_\psi$.

We parameterize the attacker with a small Transformer encoder with two layers ($d{=}256$, 4 attention heads, dropout $=0.1$), which consumes token and positional embeddings with learned positional encodings, and includes a prepended goal embedding at position $0$. The policy head maps the goal-position hidden state to logits over $\mathcal{V}$; following the implementation, we add a mild length-based logit bias favoring tokens with longer decoded strings.

Before sampling we apply a policy-side temperature to logits, mask out $\langle\mathrm{eos}\rangle$, and mask tokens outside the English-ish subset.
Let $\mathcal{A}$ denote the remaining \emph{valid} action set after masking, i.e., the support of the filtered action distribution.
We then sample actions from an $\epsilon$-mixture exploration distribution
\begin{equation}
\begin{aligned}
q(a \mid s_t, g)
&= (1-\epsilon)\,\pi_\phi(a \mid s_t, g) \\
&\quad + \epsilon\,\mathrm{Uniform}_{\mathcal{A}}(a),
\end{aligned}
\end{equation}

where $\epsilon$ decays linearly over training, as detailed in Appendix~\ref{exploration}.

\subsection{Reward Design}
\label{subsec:rl-reward}

RL-GOAL uses a terminal, length-based reward designed to favor prompts that induce extreme over-generation while discouraging trivial or degenerate prefixes (Appendix~\ref{Reward-implementation}). Let \(L\) be the continuation length produced by the victim for a given episode and \(g\) the sampled target. We form an effective goal length $g'$ by applying a minimum target $g_{\min}$ and a training-time upper cap $G_{\max}$ on the requested goal:
\begin{equation}
g' \;=\; \max\!\bigl(g_{\min},\,\min(g,\,G_{\max})\bigr),
\end{equation}
where in our implementation $g_{\min}=8192$ and $G_{\max}$ corresponds to the cap used for \texttt{max\_new\_tokens}.
We then define a normalized length ratio
\begin{equation}
\rho \;=\; \frac{L}{g'}.
\end{equation}

The primary reward term is a shaped function of the normalized length ratio $\rho$. In our implementation, the length bonus penalizes large under-shoots and saturates once the victim meets the effective goal: specifically, continuations with $\rho < 0.5$ receive a negative penalty that increases linearly as $\rho$ decreases, while for $\rho \ge 0.5$ the bonus increases linearly with $\rho$ and is clipped to a maximum value of $1$ (reaching the cap at $\rho \ge 1$). This term provides the dominant learning signal and directly incentivizes long continuations.

In addition to the length objective, the terminal reward includes (i) a permissive constant prompt-validity term, (ii) an early-EOS penalty for generating $\langle\mathrm{eos}\rangle$ substantially before the goal, and (iii) a small heuristic diversity regularizer; see Appendix~\ref{Reward-implementation} for the exact definitions and coefficients.

We combine these terms into a single scalar reward (Eq.~\ref{eq:rlgoal-reward}).

\begin{equation}
\label{eq:rlgoal-reward}
\begin{aligned}
R \;=&\; w_{\text{len}}\, b(\rho)
\;+\; w_{\text{prompt}}\, v(\mathbf{t})
\;-\; w_{\text{eos}}\, \mathrm{pen}_{\text{eos}} \\
&\;-\; \lambda\, \log\!\bigl(1 + \min(\mathrm{ppl}, 100)\bigr).
\end{aligned}
\end{equation}

where \(b(\rho)\) is the clipped length bonus, \(v(\mathbf{t})\) the prompt-validity score,
\(\mathrm{pen}_{\text{eos}}\) the early-EOS penalty, and \(\mathrm{ppl}\) a heuristic perplexity
estimate on the continuation. We initialize the reward weights (defaults: $w_{\text{len}}=0.90$, $w_{\text{prompt}}=0.10$, $w_{\text{eos}}=0.05$) and fix $\lambda=0.05$ for the perplexity regularizer.
We choose these weights so that continuation length is the primary learning signal, while auxiliary terms remain small and do not dominate optimization.
In the current implementation, the prompt-validity term is intentionally permissive (it returns $1.0$ for all prompts) and is retained mainly to preserve a stable interface for future extensions (Appendix~\ref{Reward-implementation}).
In the released trainer, the reward weights are optionally adjusted online using a simple heuristic based on recent average generation lengths (and, when available, EOS penalties); when validity scoring is disabled, the trainer skips adapting $w_{\text{prompt}}$.

\subsection{Training and Optimization}
\label{subsec:training}
We optimize \(\pi_\phi\) and \(V_\psi\) with clipped Proximal Policy Optimization (PPO) \citep{schulman-etal-2017-ppo} using trajectories collected from parallel instances of the environment (full hyperparameters in Appendix~\ref{app:ppo-hparams}). We adopt an actor--critic formulation with a learned value function $V_\psi$ as a baseline for advantage estimation, which reduces variance in policy-gradient updates and improves optimization stability \citep{sutton-etal-1999-policy,schulman2016gae,schulman-etal-2017-ppo}. Generalized advantage estimation (GAE) \citep{schulman2016gae} is used to compute targets, and we apply standard clipped policy and value losses with an entropy bonus for exploration. We maintain a prioritized replay buffer  using transition priorities proportional to absolute advantage magnitude \citep{schaul-etal-2016-prioritized} and apply hindsight goal relabeling \citep{andrychowicz2017hindsight} to a fraction of stored trajectories (Appendix~\ref{app:replay-her}). Only the compact policy and value networks are trained; the victim LLM is never fine-tuned. We prioritize replay by advantage magnitude to focus updates on high-signal transitions \citep{schaul-etal-2016-prioritized}, and use hindsight relabeling to learn from undershooting trajectories by relabeling goals with achieved lengths \citep{andrychowicz2017hindsight}. As training progresses, we expand the goal-length curriculum and increase a bootstrapping cap on the victim-side decoding budget (maximum generated tokens), moving from moderate targets to the high-OGF regime used in evaluation Section~\ref{subsec:rl-eval}. Algorithm \ref{alg:rl-goal} summarizes the training loop.

\section{Stopping-Time Metrics and Evaluation Protocol}
\label{sec:metrics}

We use a common set of attack-side stopping-time metrics to evaluate both EOGen and RL-GOAL. Let \(C\) denote the nominal context window of the victim model (in tokens), and let \(L\) be the continuation length (number of newly generated tokens) for a given query. We define the \emph{Over-Generation Factor} (OGF) as
\[
\mathrm{OGF} \;=\; \frac{L}{C}.
\]
Over-generation factors (OGF) are always reported relative to this nominal context window.

We mark a query as a \emph{stall} if it hits the generation cap without producing an EOS token, i.e., \texttt{stall}$=1$ iff \emph{no EOS} is observed and the cap is hit, and $0$ otherwise. We additionally record an \emph{over-context non-termination} flag=1 iff \emph{no EOS} is observed and \(L \ge C\).

To capture repetition in the tail, we scan the continuation with overlapping $k$-grams (with $k{=}8$), track the position of the last novel $k$-gram, and define the tail length as the number of tokens after this position. Finally, we record the wall-clock generation latency (in seconds) for each query.

\section{Experimental Setup}
\label{sec:experiments}

\paragraph{Victim Models and Context Windows}
\label{subsec:victims}

We train and evaluate EOGen on Phi-3-mini-4k-instruct \citep{abdin2024phi3technicalreporthighly}, LLaMA-2-7B-HF \citep{touvron2023llama2openfoundation}, and Deepseek-Coder-7B-Base-v1.5 \citep{guo2024deepseekcoder}, a 7B-parameter base model from the DeepSeek-Coder family. For RL-GOAL, we focus on LLaMA-2-7B-HF, as training the RL attacker is computationally intensive. Both attack methods are thus assessed on victim models.

\begin{table*}[t]
\centering
\scriptsize
\setlength{\tabcolsep}{3.9pt}
\renewcommand{\arraystretch}{1.05}
\begin{tabular}{@{}l c|rrrr|rrrr|rrrr@{}}
\toprule
\textbf{Prompt source} & \textbf{\#Prompts}
& \multicolumn{4}{c|}{ \textbf{Phi-3-mini-4k-instruct}}
& \multicolumn{4}{c|}{ \textbf{LLaMA-2-7B-HF}}
& \multicolumn{4}{c}{\textbf{Deepseek-Coder-7B-Base-v1.5}} \\
\cline{3-14}
 &  & \rule{0pt}{2.6ex}Avg.\ OGF & \rule{0pt}{2.6ex}S@$\ge$1 & \rule{0pt}{2.6ex}S@$\ge$2 & \rule{0pt}{2.6ex}S@$\ge$4
    & \rule{0pt}{2.6ex}Avg.\ OGF & \rule{0pt}{2.6ex}S@$\ge$1 & \rule{0pt}{2.6ex}S@$\ge$2 & \rule{0pt}{2.6ex}S@$\ge$4
    & \rule{0pt}{2.6ex}Avg.\ OGF & \rule{0pt}{2.6ex}S@$\ge$1 & \rule{0pt}{2.6ex}S@$\ge$2 & \rule{0pt}{2.6ex}S@$\ge$4 \\

\midrule

EOGen        & 871/863/524     & 1.39$\pm$1.14 & 52.1\% & 25.2\% & 6.8\%
                          & 0.47$\pm$0.68 & 17.4\% & 4.5\%  & 0.7\%
                          & 0.49$\pm$0.87 & 13.3\% & 7.7\%  & 1.6\% \\
EOGen-suffix & 871/863/524     & 0.74$\pm$0.85 & 25.3\% & 8.2\%  & 1.7\%
                          & 0.58$\pm$0.60 & 25.4\% & 3.2\%  & 0.1\%
                          & 0.59$\pm$0.75 & 23.7\% & 6.8\%  & 0.0\% \\
\midrule
\multicolumn{14}{l}{\textbf{Baselines}} \\
\midrule
Repeat-style & 25/25/25        & 0.54$\pm$0.60 & 14.0\% & 3.8\%  & 0.4\%
                          & 0.32$\pm$0.45 & 11.5\% & 0.9\%  & 0.1\%
                          & 0.27$\pm$0.48 & 9.4\%  & 1.6\%  & 0.0\% \\
Inf.\ babble & 30/30/30        & 0.96$\pm$0.74 & 33.3\% & 9.4\%  & 0.5\%
                          & 0.40$\pm$0.53 & 12.2\% & 2.1\%  & 0.3\%
                          & 0.19$\pm$0.39 & 5.7\%  & 0.9\%  & 0.0\% \\
Random short & 100/100/100     & 0.51$\pm$0.72 & 13.9\% & 5.2\%  & 1.3\%
                          & 0.42$\pm$0.67 & 14.0\% & 4.6\%  & 0.7\%
                          & 0.50$\pm$0.80 & 16.3\% & 5.8\%  & 1.8\% \\
WizardLM     & 100/100/100     & 0.27$\pm$0.37 & 4.0\%  & 1.0\%  & 0.1\%
                          & 0.23$\pm$0.31 & 4.7\%  & 0.2\%  & 0.0\%
                          & 0.24$\pm$0.43 & 5.4\%  & 1.2\%  & 0.2\% \\
\bottomrule
\end{tabular}
\caption{Each prompt is evaluated with a fixed budget $B{=}4C$ for each victim model. \#Prompts is reported as Phi-3 / LLaMA-2 / DeepSeek-Coder.}
\label{tab:eogen-vs-baselines-all}
\end{table*}

\section{Experimental Results}
\subsection{EOGen}
\label{sec:results}

\paragraph{Prompt set and evaluation harness.}
(i) EOGen We evaluate the saved prompt set from all three models where; each prompt is evaluated with $T{=}10$ stochastic continuations. (ii) For EOGen-suffix, we evaluate the same saved prompts under identical decoding settings by appending each prompt as a suffix to a WizardLM instruction (i.e., \texttt{instruction},$+$,\texttt{prompt}).
All trials are executed via the victim model's \texttt{generate} interface with
\texttt{do\_sample=True}, \texttt{temperature=1.0}, and \texttt{top\_p=1.0}, using a fixed cap of
$B{=}4C$ new tokens for both tests. The nominal context window $C$ is read from the model configuration. 

\paragraph{Baselines.}
To contextualize EOGen, we evaluate baseline prompt families under the same victim access and sampling
configuration as above. These include: (i) hand-crafted repeat-style prompts
(\texttt{repeat}, \texttt{recursion}, \texttt{count}, \texttt{longtext}, \texttt{code}),
(ii) an \texttt{infinite babble} prompt that explicitly requests non-termination, (iii) \texttt{random short} prefixes consisting of 3--7 words sampled from a fixed word pool, and (iv) a WizardLM-style instruction baseline when a WizardLM dataset \citep{xu-etal-2024-wizardlm} is provided. 
For baselines we set the generation budget to $B{=}4C$ and evaluate each prompt with 34 stochastic trials under the same budget to approximately match the total number of sampled victim continuations to EOGen (about $871\times10\approx8.7$k), yielding a comparable overall sample count when baselines use 25--100 prompts. Success@OGF$\ge 1$ is implemented as $L \ge C$, where $L$ is the number of newly generated tokens.

\paragraph{Main Results}
\label{subsec:main-results}
Table~\ref{tab:eogen-vs-baselines-all} compares EOGen and EOGen-suffix against baseline
prompt families across three victim models. Figure~\ref{fig:sample_EOGen_suffix} provides a representative qualitative example for EOGen-suffix. Across victims, EOGen increases over-generation severity relative to all baselines, yielding higher
Success@OGF thresholds (S@$\ge$1/2/4) and larger mean OGF.
In contrast, the random-short control achieves comparatively low Success@OGF at the same budget,
indicating that severe over-generation is uncommon under short random prefixes.
Latency results for EOGen are reported in Appendix~\ref{app:C2-latency}.

\paragraph{Prompt-level robustness and stalling.}
\label{subsec:prompt-level}

\begin{figure*}[t]
  \centering
  \begin{subfigure}[t]{0.49\textwidth}
    \centering
    \includegraphics[width=\linewidth]{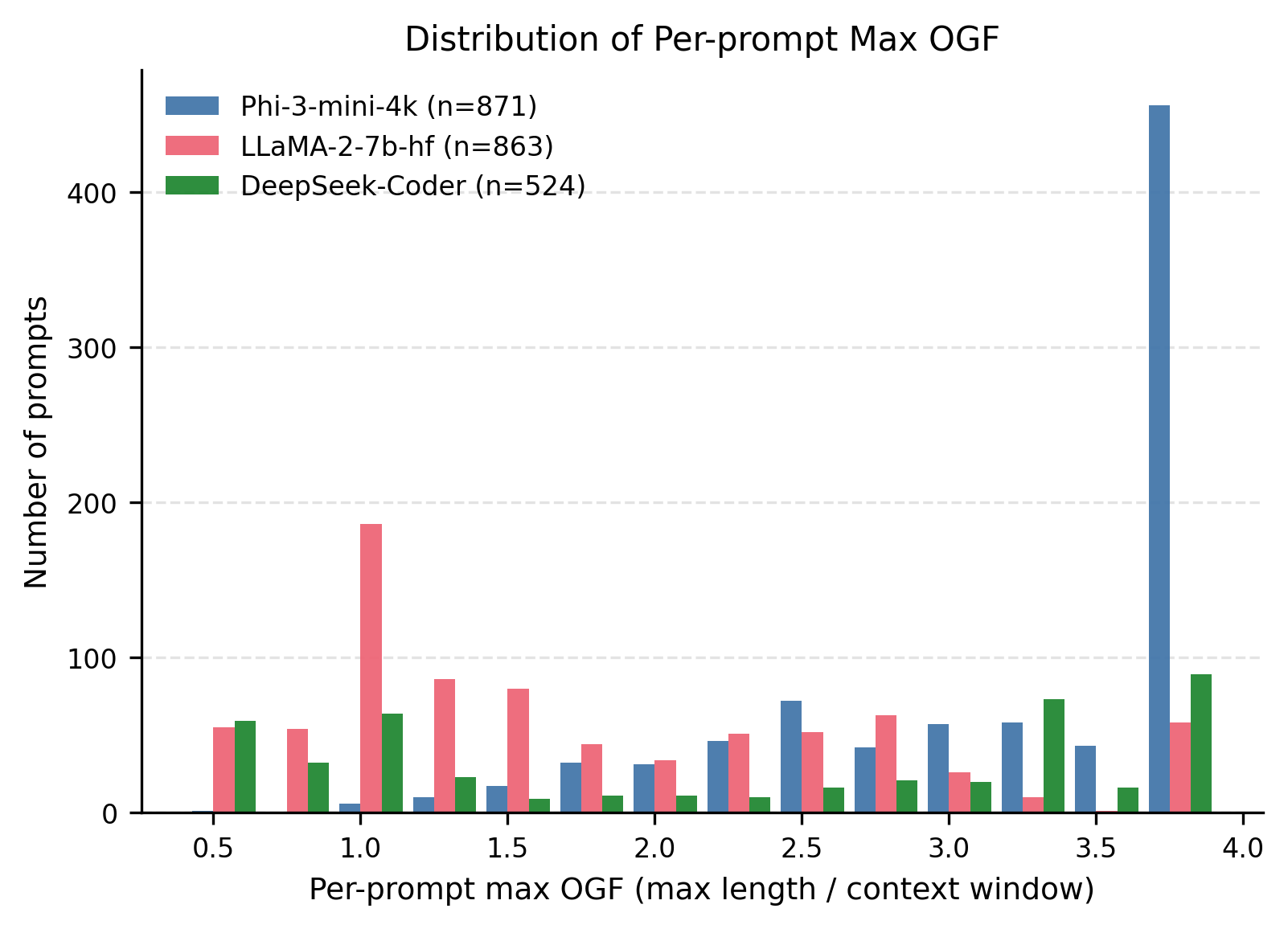}
    \caption{Histogram of per-prompt maximum OGF across three victim models.}
    \label{fig:eogen-succ-dist}
  \end{subfigure}
  \hfill
  \begin{subfigure}[t]{0.49\textwidth}
    \centering
    \includegraphics[width=\linewidth]{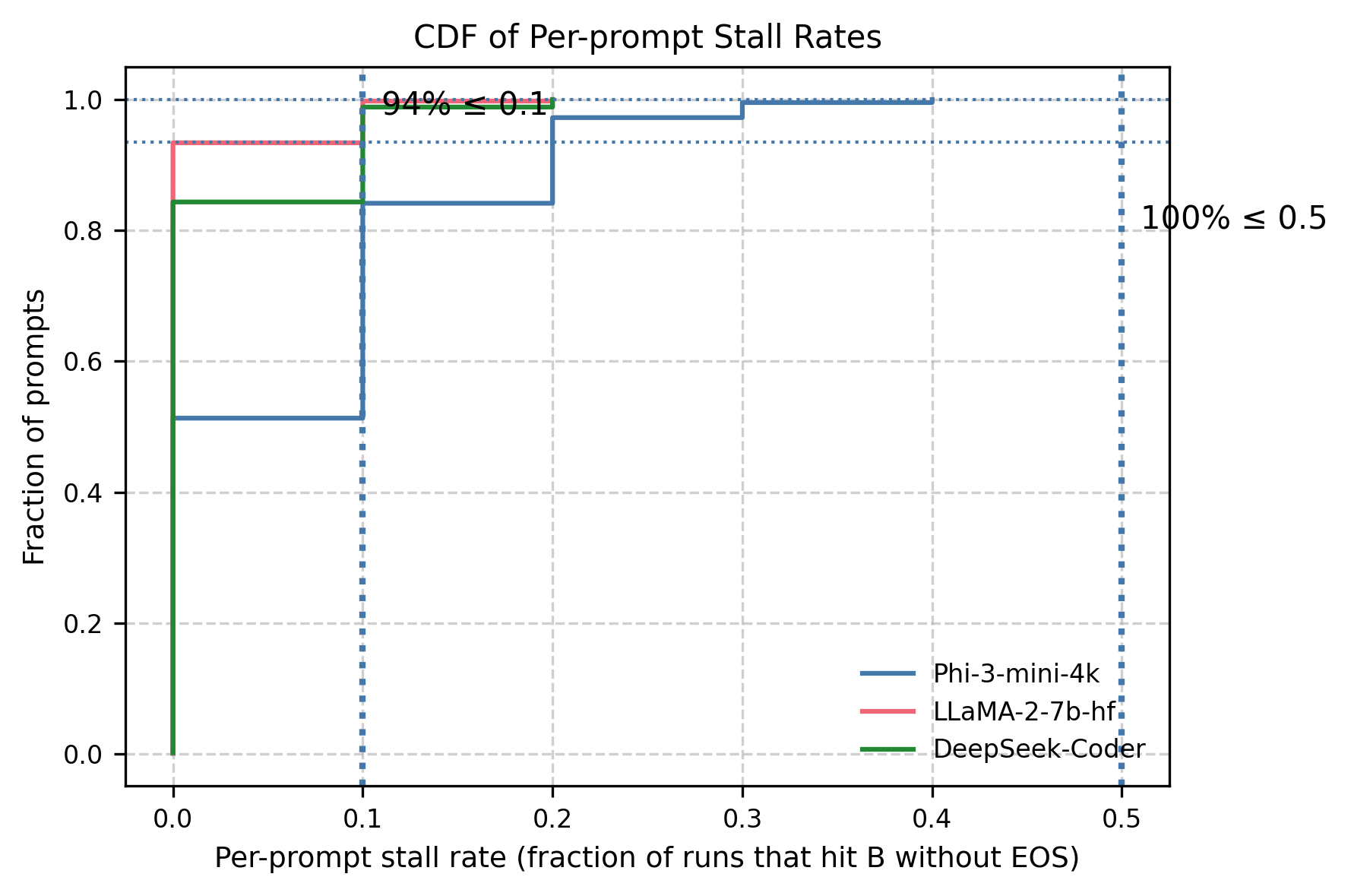}
    \caption{CDF of per-prompt stall rates (fraction of trials that reach $B$ without $\langle\mathrm{eos}\rangle$).}
    \label{fig:eogen-stallcdf}
  \end{subfigure}
  \caption{Prompt-level robustness and non-termination for EOGen.}
  \label{fig:eogen-prompt-level}
\end{figure*}

Aggregate success rates can hide substantial variation across prompts.
To characterize robustness at the \emph{prompt level}, we compute per-prompt over-generation and stall rates across the $T{=}10$ stochastic trials.
Figure~\ref{fig:eogen-prompt-level} summarizes these distributions: panel~(a) shows a histogram of the \emph{per-prompt maximum OGF} (max over trials), highlighting how strongly different prompts can trigger over-generation under sampling, while panel~(b) plots the CDF of \emph{per-prompt stall rates} (fraction of trials that reach $B$ without $\langle\mathrm{eos}\rangle$), revealing that non-termination is concentrated in a small tail of prompts rather than being uniform across the prompt set.
Together, these plots expose substantial heterogeneity in prompt robustness and a heavy-tailed pattern of stalling.

\begin{table*}[t]
\centering
\small
\setlength{\tabcolsep}{5.0pt}
\begin{tabular}{lccccccc}
\toprule
\textbf{Setting} & \textbf{Avg.\ OGF} & \textbf{Succ.@$\ge 1$} & \textbf{Succ.@$\ge 2$} & \textbf{Succ.@$\ge 4$} & \textbf{Stall} & \textbf{Avg.\ $L$} & \textbf{Latency (s)} \\
\midrule
\multicolumn{8}{l}{\textbf{RL-GOAL (trained policy)}} \\
\midrule
Llama-2-7b-hf            & $2.04\pm1.37$ & 88.2\% & 36.0\% & 28.7\% & 28.7\% & $8342\pm5608$  & $288.9\pm244.3$ \\
Llama-2-13b-chat-hf      & $0.50\pm0.51$ & 35.2\% & 0.1\%  & 0.0\%  & 0.0\%  & $2055\pm2075$  & $52.5\pm54.3$   \\
Phi-3-mini-4k-instruct   & $2.70\pm1.43$ & 79.4\% & 64.3\% & 46.0\% & 46.0\% & $11061\pm5855$ & $220.2\pm121.7$ \\
EleutherAI/pythia-6.9b   & $1.52\pm1.28$ & 72.9\% & 13.6\% & 7.6\%  & 0.3\%  & $3113\pm2630$  & $82.5\pm89.2$   \\
\addlinespace
\midrule
\multicolumn{8}{l}{\textbf{Baseline (No Prefix)}} \\
\midrule
Llama-2-7b-hf            & $0.36\pm0.58$ & 12.8\% & 3.1\%  & 0.2\%  & 0.2\%  & $1477\pm2356$  & $39.8\pm74.4$   \\
Llama-2-13b-chat-hf      & $0.09\pm0.13$ & 0.5\%  & 0.0\%  & 0.0\%  & 0.0\%  & $369\pm513$    & $9.0\pm14.6$    \\
Phi-3-mini-4k-instruct   & $0.66\pm0.82$ & 16.4\% & 7.8\%  & 2.3\%  & 2.3\%  & $2687\pm3343$  & $48.7\pm68.7$   \\
EleutherAI/pythia-6.9b   & $0.82\pm1.54$ & 20.8\% & 10.9\% & 4.9\%  & 1.8\%  & $1671\pm3155$  & $49.3\pm113.4$  \\
\bottomrule
\end{tabular}
\caption{Main results (mean$\pm$std over trials). Success@$\ge k$ denotes the fraction of prompts achieving $\mathrm{OGF}\ge k$. Stall is the fraction of trials that reach budget $B$ without emitting \texttt{<eos>}.}
\label{tab:rlgoal-main}
\end{table*}

\subsection{RL Evaluation Protocol}
\label{subsec:rl-eval}

At test time we freeze the trained RL-GOAL attacker and use it solely as a prefix generator in a generate-only / query-only interface setting. Each evaluation run conditions the attacker on a fixed target length $B$ (we use $B{=}16,384$); in our evaluation we set the target length equal to the decoding cap, i.e., $B{=}16,384$, and we sample an attacker prefix for up to $T$ steps (default $T{=}64$). The attacker is queried on a trailing observation window of 1024 tokens, and actions are sampled using temperature sampling with nucleus filtering (default $\tau{=}0.8$, $p{=}0.95$). During prefix rollout, EOS is not masked by default; instead, the harness exposes an optional EOS gating mechanism that can defer EOS until after a fraction of the length cap, but this gating is disabled in our reported runs.

The victim model is queried exactly once per run: we concatenate the benign base prompt with the attacker prefix and trigger a single victim continuation under the evaluation caps used by the harness. We then compute length-based stopping-time statistics from the victim continuation length $L$, including the over-generation factor $\mathrm{OGF}{=}L/C$ for nominal context window $C$, Success@$\mathrm{OGF}\ge\{1,2,3,4\}$, and a cap-hit indicator (\texttt{stall}) defined by reaching the configured new-token cap. We also report wall-clock latency measured around the victim-side generation call. (Token-level tail statistics are not computed in this harness because only continuation lengths are logged.)

\subsection{Main results}
\label{subsec:rlgoal-main}

RL-GOAL is trained using LLaMA-2 as the victim, and therefore constructs prefixes in the LLaMA token space. When evaluating against a non-LLaMA victim , we bridge tokenizers by decoding the generated prefix with the LLaMA tokenizer and re-encoding it with the victim tokenizer before invoking the victim’s \texttt{generate} call, ensuring the victim receives correctly tokenized inputs under its own vocabulary.

Table~\ref{tab:rlgoal-main} summarizes 1000 evaluation runs per victim model, compared to a no-prefix baseline under the same victim decoding configuration. RL-GOAL induces strong over-generation on LLaMA-2-7B and Phi-3: average OGF reaches $2.04\pm1.37$ and $2.70\pm1.43$, with Success@$\ge 2$ of 36.0\% and 64.3\% (baseline: 3.1\% and 7.8\%). Latency distributions and amplification are shown in Appendix~\ref{app:C3-latency} (Figure~\ref{fig:rlgoal-latency}). These gains are accompanied by frequent stalls (28.7\% / 46.0\%), i.e., cap-saturating generations without EOS, indicating that the learned prefixes often drive decoding to the configured limit. LLaMA-2-13B-Chat is comparatively robust: while RL-GOAL improves Success$\ge1$ (35.2\% vs.\ 0.5\%), Success@$\ge 4$ remains 0.0\% and stalls remain 0.0\%. Pythia shows moderate susceptibility (Avg.\ OGF $1.52\pm1.28$, Success@$\ge4$ 7.6\%) with rare stalls (0.3\%). Latency increases with continuation length, with the highest mean runtimes observed for the models with the largest average $L$ (LLaMA-2-7B, Phi-3). A visual summary of these trial-level aggregates is provided in Appendix~\ref{app:C4-RL-GOAL summary metrics} (Figure~\ref{fig:rlgoal-overview}).

\section{Conclusion}
We introduced EOGen and RL-GOAL, two black-box prompt-based attack methods that induce sustained over-generation, amplifying inference cost and creating a practical denial-of-service mechanism in large language models. Across multiple open-source victims, we show that simple evolutionary search can already find effective short prompts, while RL-GOAL further increases attack severity by learning a policy that reliably drives long, budget-hitting continuations. These results highlight that over-generation is a measurable and reproducible failure mode under standard decoding interfaces, and motivate stronger deployment-time safeguards for future LLM systems.

\section{Limitations}
\label{sec:limitations}

EOGen and RL-GOAL are heuristic optimizers and therefore do not explore the full combinatorial prompt space. For tractability, we restrict the search (e.g., short prompts, bounded editing steps, limited token subsets); RL-GOAL further narrows exploration via action constraints (e.g., ``English-like'' token masking and disallowing EOS actions during policy sampling), which may miss effective but out-of-scope attacks. RL-GOAL is also substantially more expensive than EOGen: PPO training requires many victim rollouts under long decoding budgets, increasing wall-clock time and inference cost. While a curriculum that ramps the generation cap improves stability, it can bias exploration toward behaviors reachable under smaller early budgets and further increases training time and total query usage.

Finally, both methods assume consistent access to the victim \texttt{generate} interface and decoding controls; in deployed APIs, rate limits, filtering, throttling, or provider-side changes can introduce non-stationarity and reduce reproducibility. Our objective is intentionally length-dominant, not plausibility- or stealth-oriented, so discovered prompts may be unnatural and easier to detect than attacks optimized for fluency, and we do not train an explicit acceptability model. Empirically, results are measured on a limited set of (relatively small) open-source victims and a fixed budgeting convention; generalization to larger and longer-context models, alternative decoding policies (e.g., stop sequences, repetition penalties), or different prompting formats remains to be tested.

In particular, we do not evaluate larger-scale frontier or long-context models (e.g., $\ge$30B parameters or $\ge$32K context), so the extent to which our findings scale with model size, context length, or provider-side guardrails is unknown.

\section*{Ethics considerations}
We study denial-of-service style over-generation vulnerabilities in LLMs to support safer deployment.
We do not release prompts or configurations that materially increase misuse risk; we report aggregate findings
and focus on mitigation and evaluation practices. We follow responsible disclosure norms where applicable.


\bibliography{custom}

\appendix

\begin{table*}[t]
\centering
\scriptsize
\setlength{\tabcolsep}{7.0pt}
\renewcommand{\arraystretch}{1.05}
\begin{tabular}{@{}l|rrrr|rrrr|rrrr@{}}
\toprule
\textbf{Prompt source}
& \multicolumn{4}{c|}{\textbf{Phi-3-mini-4k-instruct}}
& \multicolumn{4}{c|}{\textbf{LLaMA-2-7B-HF}}
& \multicolumn{4}{c}{\textbf{Deepseek-Coder-7B-Base-v1.5}} \\
\cline{2-13}
& \rule{0pt}{2.6ex}\textbf{Median} & \textbf{Mean} & \textbf{P95} & \textbf{Max}
& \rule{0pt}{2.6ex}\textbf{Median} & \textbf{Mean} & \textbf{P95} & \textbf{Max}
& \rule{0pt}{2.6ex}\textbf{Median} & \textbf{Mean} & \textbf{P95} & \textbf{Max} \\
\midrule
EOGen        & 78.17 & 108.45 & 329.86 & 332.66  & 17.03 & 53.19 & 221.14 & 652.68  & 13.91 & 59.13 & 419.74 & 628.60 \\
EOGen-suffix & 23.57 & 55.27  & 218.28 & 333.82  & 32.67 & 62.54 & 184.50 & 653.31  & 22.54 & 65.31 & 289.90 & 535.21 \\
\midrule
\multicolumn{13}{l}{\textbf{Baselines}} \\
\midrule
Repeat-style & 21.09 & 37.72  & 137.04 & 323.28  & 11.93 & 33.36 & 128.67 & 649.11  & 6.44  & 28.00 & 118.91 & 502.52 \\
Inf.\ babble & 49.28 & 70.42  & 200.49 & 329.59  & 17.94 & 42.39 & 147.69 & 648.05  & 4.43  & 19.71 & 106.25 & 458.71 \\
Random short & 14.03 & 36.12  & 160.47 & 332.40  & 12.52 & 47.16 & 226.76 & 650.15  & 15.75 & 57.32 & 304.31 & 625.65 \\
WizardLM     & 9.27  & 17.89  & 63.70  & 331.52  & 10.28 & 22.33 & 97.93  & 412.81  & 9.49  & 24.76 & 104.89 & 630.50 \\
\bottomrule
\end{tabular}
\caption{Generation latency (seconds), measured as wall-clock time. Columns report median/mean/95th percentile/max latency.}
\label{tab:latency-phi3}
\end{table*}

\section{EOGen - Implementation Details}

\subsection{Hyperparameters}
\label{subsec:eogen-hparams}

Throughout EOGen, we restrict candidate prompts to natural, human-readable English (ASCII alphabetic tokens present in an English word list) to reflect a realistic query-only attacker who prefers deployable prompts over non-linguistic character artifacts.

For EOGen we use a small but fixed evolutionary budget for all victims. We set the population size to \(P = 20\), the selection fraction to \(\rho = 0.5\) (truncation selection on the top half of the population), and the token-wise mutation rate to \(\mu = 0.05\). New candidates are generated by repeatedly sampling parent pairs from the survivor set, applying one-point crossover, then performing independent per-token mutation with rate $\mu$. We enforce the length constraint $[\ell_{\min}, \ell_{\max}]$ by padding with randomly sampled allowed tokens or truncating as needed. These hyperparameters follow common practice in evolutionary search, balancing exploitation (selecting the top-performing half of the population) with exploration (token-wise mutation), while keeping the total query budget tractable \citep{guo-etal-2025-evoprompt}. Prompts are constrained to have length \(\ell \in [\ell_{\min}, \ell_{\max}]\) with \(\ell_{\min} = 3\) and \(\ell_{\max} = 7\) tokens (as described in Section~\ref{subsec:prompt-space}). We run EOGen for $G{=}500$ generations, yielding approximately $P\times G \approx 10^{4}$ victim queries per run (with $P{=}20$). We use the same configuration for LLaMA-2 and Phi-3; for DeepSeek-Coder we ran fewer generations ($G{=}387$) due to limited compute time, resulting in a smaller query budget for that victim. We report DeepSeek-Coder results with this reduced budget.

\subsection{Decoding Configuration}
\label{subsec:decoding}

For all attacks we treat the victim as a black‑box text‑completion API with a fixed decoding regime. We enable sampling via \texttt{do\_sample} but do not set explicit temperature, nucleus top-$p$ or repetition penalty , so generation uses the library defaults for these parameters. We cap newly generated tokens at $B=4C$, where $C$ is the victim’s nominal context window. This cap is fixed across all prompts and generations for a given victim.

\subsection{Implementation Notes}
\label{subsec:notes}
Our released code supports checkpoint-based resume by saving (and restoring) the current generation index, population, and best-so-far prompt/reward at the end of each generation. It also logs one row per prompt evaluation, including the fitness value and diagnostic fields such as continuation length, OGF, stall flag, tail-persistence, and latency. During search, the implementation maintains running \texttt{Success@OGF$\geq$2} and \texttt{Success@OGF$\geq$4} counts as a function of total evaluations and periodically saves a budgeted-success curve plot; additionally, it archives prompts whose continuations exceed 8192 tokens and tracks the best prompt that exactly reaches the generation cap ($B=16{,}384$).

\label{subsec:EOAlgorithm}
\begin{algorithm}[t]
\caption{EOGen: Evolutionary Over-Generation Prompt Search}
\label{alg:eogen}
\begin{algorithmic}[1]\small
\STATE \textbf{Input:} victim model $f_\theta$, allowed token set $\mathcal{S}$, length bounds $[\ell_{\min},\ell_{\max}]$, population size $P$, generations $G$
\STATE \textbf{Hyperparameters:} fitness $R(\cdot)$, mutation rate $\mu$, selection fraction $\rho$
\STATE \textbf{Output:} best prompt $\mathbf{t}^\star$
\STATE Initialise $P^{(0)}=\{\mathbf{t}^{(1)},\dots,\mathbf{t}^{(P)}\}$ by seeding from ``In'' and padding/truncating with tokens from $\mathcal{S}$ to satisfy $[\ell_{\min},\ell_{\max}]$
\STATE $\mathbf{t}^\star \leftarrow \texttt{None}$;\quad $R^\star \leftarrow -\infty$
\FOR{$g = 0,1,\dots,G-1$}
  \FOR{each $\mathbf{t} \in P^{(g)}$}
    \STATE $\mathbf{y}\leftarrow \textsc{Generate}(f_\theta,\mathbf{t})$
    \STATE $R(\mathbf{t})\leftarrow \textsc{Fitness}(\mathbf{t},\mathbf{y})$
    \IF{$R(\mathbf{t})>R^\star$}
      \STATE $\mathbf{t}^\star\leftarrow \mathbf{t}$;\quad $R^\star\leftarrow R(\mathbf{t})$
    \ENDIF
  \ENDFOR
  \STATE $S^{(g)} \leftarrow$ top $\rho P$ prompts in $P^{(g)}$ ranked by $R(\mathbf{t})$ \COMMENT{$\rho=0.5$}
  \STATE $P^{(g+1)} \leftarrow S^{(g)}$
  \WHILE{$|P^{(g+1)}| < P$}
    \STATE Sample two distinct parents $\mathbf{t}^{(1)},\mathbf{t}^{(2)}$ uniformly from $S^{(g)}$
    \STATE $\mathbf{u} \leftarrow \textsc{OnePointCrossover}(\mathbf{t}^{(1)},\mathbf{t}^{(2)})$
    \STATE $\mathbf{u} \leftarrow \textsc{Mutate}(\mathbf{u},\mu)$
    \STATE $\mathbf{u} \leftarrow \textsc{ProjectLen}(\mathbf{u},[\ell_{\min},\ell_{\max}],\mathcal{S})$
    \STATE Add $\mathbf{u}$ to $P^{(g+1)}$
  \ENDWHILE
\ENDFOR
\STATE \textbf{return} $\mathbf{t}^\star$
\end{algorithmic}
\end{algorithm}

\section{RLGOAL - Implementation Details}

\subsection{Prompt initialization.}
\label{app:prompt-init}
During training, we mix a fixed benign header with base prompts sampled from a prompt pool mined from prior episodes.
The probability of using the fixed header decays over the first $0.2\times$ training episodes and is floored at $0.10$.
When sampling from the prompt pool, prompts are drawn with probability proportional to their stored score (used as a sampling weight).

\subsection{Environment and episode structure.}
\label{subsec:rl-impl}
RL-GOAL is implemented as a Gym-style environment in which the agent appends one token per step to a prompt prefix and receives a \emph{terminal} reward at episode end. Episodes run for at most a fixed horizon ($T{=}64$) to restrict prefix length and control training cost, since each step triggers a victim call. During training, the EOS action is masked in the policy sampling distribution. We implement two overshoot safeguards based on an \emph{agent-budget} $g_{\text{agent}}=\min(g,B_{\max})$: (i) a trainer-side early stopping rule that ends an episode once the agent body exceeds $1.2\times g_{\text{agent}}$, and 
(ii) an environment-side hard cutoff that appends EOS and terminates once the body exceeds $1.3\times g_{\text{agent}}$.
Intermediate shaping is disabled (per-step reward is zero), so learning is driven by the terminal signal.

A token is considered valid if its decoded string is mostly composed of common ASCII characters (letters/digits/space/punctuation), as determined by a lightweight heuristic; EOS is handled separately by explicit masking during training-time sampling.

The released training entry-point executes a short smoke-test loop prior to training; it interacts with the victim only through the same terminal \texttt{generate}-based reward query as training episodes.

\subsection{Goal curriculum and bootstrapping cap.}
\label{app:rl-curriculum}
We sample one goal $g$ per training batch from a moving interval whose upper bound increases over training, while respecting $G_{\max}$ and the current bootstrapping cap ($b_k$).
To emphasize long-target behavior, we bias sampling toward the upper half of the interval, increasing this bias once the boot cap reaches the long-context regime.
The boot cap increases in stages (starting at 2048 and incrementing by 2048 every 5000 episodes up to 16384), throttling the victim's decoding budget early in training.

\paragraph{Curriculum cap and its interaction with bootstrapping.}
The curriculum cap $c_k$ is a monotone schedule that expands the \emph{support} of the goal sampler over training by raising the largest admissible targets, while the boot cap $b_k$ constrains feasibility by limiting the victim’s available decoding budget early in training. As a result, when $b_k$ is small the curriculum is effectively truncated by the boot cap, and increasing $c_k$ has no effect until $b_k$ grows; once $b_k$ enters the long-context regime, $c_k$ becomes the active limiter and the sampler’s mass shift toward the upper half of the interval increases accordingly. In this way, $c_k$ controls \emph{which} long targets are requested as training progresses, whereas $b_k$ controls \emph{when} those targets become attainable by the victim under the fixed generation cap.

\subsection{Victim query.}
\label{app:rl-victim-query}
At episode end, we query the victim LLM exactly once using \texttt{generate} with a per-episode decoding budget $B$ equal to the minimum of $g$, the global generation cap, and the current bootstrapping cap.
We use stochastic decoding (\texttt{do\_sample}=\texttt{True}, temperature $=0.9$, top-$p$ $=0.95$, repetition penalty $=1.05$) and include a logits processor that replaces NaN/Inf logits with finite values for numerical stability. The continuation length $L$ and EOS position (if any) are extracted from the returned sequence and used to compute the terminal reward.

\subsection{Policy sampling and exploration.}
\label{exploration}
During rollouts, we sample actions from the filtered token distribution using a policy temperature of $0.8$ and an $\epsilon$-mixture exploration strategy, where $\epsilon$ decays linearly from $0.01$ to $0.001$ over 20{,}000 training episodes.
Following the implementation, we add a small length-based logit bias (coefficient $0.4$) proportional to each token's decoded string length to accelerate exploration over discrete token choices. We use a compact 2-layer Transformer to keep attacker overhead negligible relative to victim-query cost and to stabilize learning under a fixed query budget.

\subsection{Reward implementation.}
\label{Reward-implementation}
At episode end, the environment computes a \emph{terminal} reward from the \emph{victim continuation} by
evaluating the full prompt (seed prefix + agent tokens) with a fresh victim free-run decoding
(\texttt{generate}) and extracting the continuation length $L$ and EOS position.
The reward depends on $L$ relative to an effective goal $g'=\max\!\bigl(8192,\,\min(g,\,G_{\max})\bigr)$, where $G_{\max}$ is the configured global new-token cap used during training; thus, all rewards are normalized with respect to a minimum target of 8192 tokens, even if the sampled goal $g$ is shorter. We introduce the $8192$-token floor to focus learning on the long-context failure mode (extreme over-generation) and to keep reward magnitudes comparable when shorter sampled goals would otherwise dominate early training.

For $\rho = L/g'$, we apply the same length-based shaping described in
Section~\ref{subsec:rl-reward}: episodes with $\rho < 0.5$ incur a penalty, whereas those with $\rho$
approaching 1 (or beyond) earn a bonus up to a maximum of 1.0.
We use a permissive prompt-validity score that returns $1.0$ for all prompts and apply an early-EOS
penalty only when the first EOS occurs before $0.9\,g$.
Finally, we include a mild heuristic regularizer derived from the continuation’s distinct-token statistics (a proxy quantity, not true perplexity), by subtracting $0.05\log(1+\min(\mathrm{ppl},100))$.

\subsection{Optimization (PPO hyperparameters).}
\label{app:ppo-hparams}
We train the policy and value networks with PPO using discount $\gamma{=}0.995$ and GAE parameter $\lambda{=}0.9$.
We use PPO clipping $\epsilon_{\text{clip}}{=}0.2$, entropy coefficient $0.005$, and value loss coefficient $0.1$.
Both networks are optimized with Adam (learning rate $2\times 10^{-5}$), and gradients are clipped to a maximum norm of $1.0$.
Each PPO update runs $6$ epochs, iterating over mini-batches of size $4$.

\subsection{Replay, prioritization, and HER.}
\label{app:replay-her}
We store transitions in a replay buffer of capacity $75{,}000$ with prioritized sampling; priorities are proportional to the magnitude of the (normalized) advantage.
We apply hindsight experience relabeling to a fraction of stored transitions: the achieved length (clipped to a maximum of $16{,}384$ tokens) replaces the original goal.
The relabeled reward is $1.0$ if the achieved length is within $\pm 5$ tokens of the relabeled goal; otherwise it is a negative distance-based shaping term.
The HER fraction is initialized to $0.3$ and adaptively adjusted based on the recent success rate.

\begin{algorithm}[t]
\small
\caption{RL-GOAL: Goal-Conditioned Reinforcement Learning Attacker}
\label{alg:rl-goal}
\textbf{Input:} Black-box victim $f_\theta$, environment $\mathcal{E}$, policy $\pi_\phi$, value $V_\psi$ \\
\textbf{Hyperparameters:} horizon $T$, victim generation caps (\texttt{MAX\_NEW\_GEN}, \texttt{BOOT\_CAP}), goal curriculum, PPO updates \\
\textbf{Output:} Trained goal-conditioned attacker $\pi_\phi$

\begin{algorithmic}[1]
\STATE Initialize policy and value parameters $\phi, \psi$
\FOR{training iteration $k = 1, 2, \dots$}
  \STATE \textbf{Collect rollouts}
  \FOR{each parallel environment instance}
    \STATE Sample target length $g \sim \textsc{GoalCurriculum}(k)$
    \STATE Reset environment: $(s_0, g) \leftarrow \mathcal{E}.\textsc{Reset}(g)$
    \FOR{$t = 0$ to $T-1$}
      \STATE Sample action $a_t \sim \pi_\phi(\cdot \mid s_t, g)$ over filtered tokens
      \STATE Step environment: $(s_{t+1}, r_t, \textit{done}, \textit{info}) \leftarrow \mathcal{E}.\textsc{Step}(a_t)$
      \STATE Store $(s_t, g, a_t, r_t, \textit{done}, V_\psi(s_t,g), \log \pi_\phi(a_t \mid s_t,g))$
      \IF{\textit{done}}
        \STATE \textbf{break}
      \ENDIF
    \ENDFOR
  \ENDFOR
  \STATE Push collected trajectories into a replay buffer (with priorities based on normalized advantage)
  \STATE Optionally apply HER by relabeling a fraction of trajectories with achieved goals
  \STATE \textbf{Compute returns and advantages} using $V_\psi$ and GAE
  \STATE \textbf{PPO update (from replay samples)}
  \FOR{epoch $e = 1$ to $E$}
    \STATE Sample mini-batches from the replay buffer
    \STATE Update $\phi$ with clipped policy loss + entropy bonus
    \STATE Update $\psi$ with clipped value loss
  \ENDFOR
  \STATE Optionally update goal curriculum and victim caps (\texttt{MAX\_NEW\_GEN}, \texttt{BOOT\_CAP})

\ENDFOR
\STATE \textbf{return} trained policy $\pi_\phi$
\end{algorithmic}
\end{algorithm}

\section{Additional Metrics}    
\subsection{Additional Prompt-level Severity}
\label{app:extra-metrics}

\begin{figure}[!t]
  \centering
  \includegraphics[width=\columnwidth]{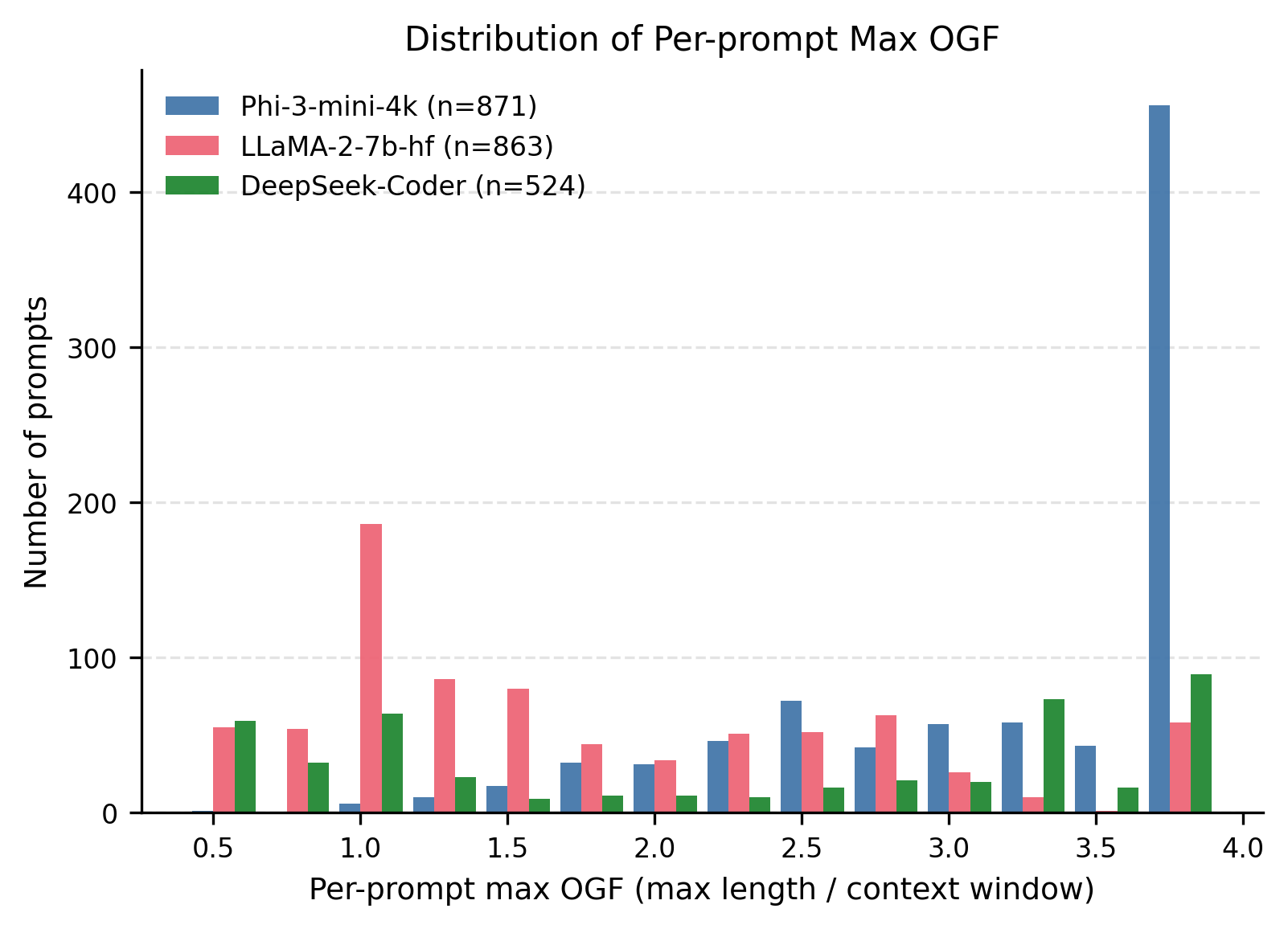}
  \caption{Histogram of per-prompt max OGF (max over $T{=}10$ trials).Max-over-trials is an optimistic summary; per-trial stall (reaching $B$ without \texttt{<eos>}) is in Table~\ref{tab:eogen-vs-baselines-all}.}
  \label{fig:eogen-maxogf}
\end{figure}

We report EOS rate and stall rate in Figure~\ref{fig:eogen-maxogf}; these are closely tied to cap-hit events under the fixed budget $B$ and are largely redundant with per-trial Success@OGF$\ge 4$ in our setting ($B=4C$).

\subsection{Latency - EOGen}
\label{app:C2-latency}
Latency reported in Table~\ref{tab:latency-phi3} is measured as wall-clock time. We parallelize evaluation via job arrays using 15 chunks, so that each job evaluates a disjoint subset of prompts. Each chunk is run as single-process inference on one NVIDIA L40S GPU; reported latencies are per-trial wall-clock times.

\subsection{Latency - RLGOAL}
\label{app:C3-latency}
\begin{figure*}[t]
  \centering
  \begin{subfigure}[t]{0.48\textwidth}
    \centering
    \includegraphics[width=\linewidth]{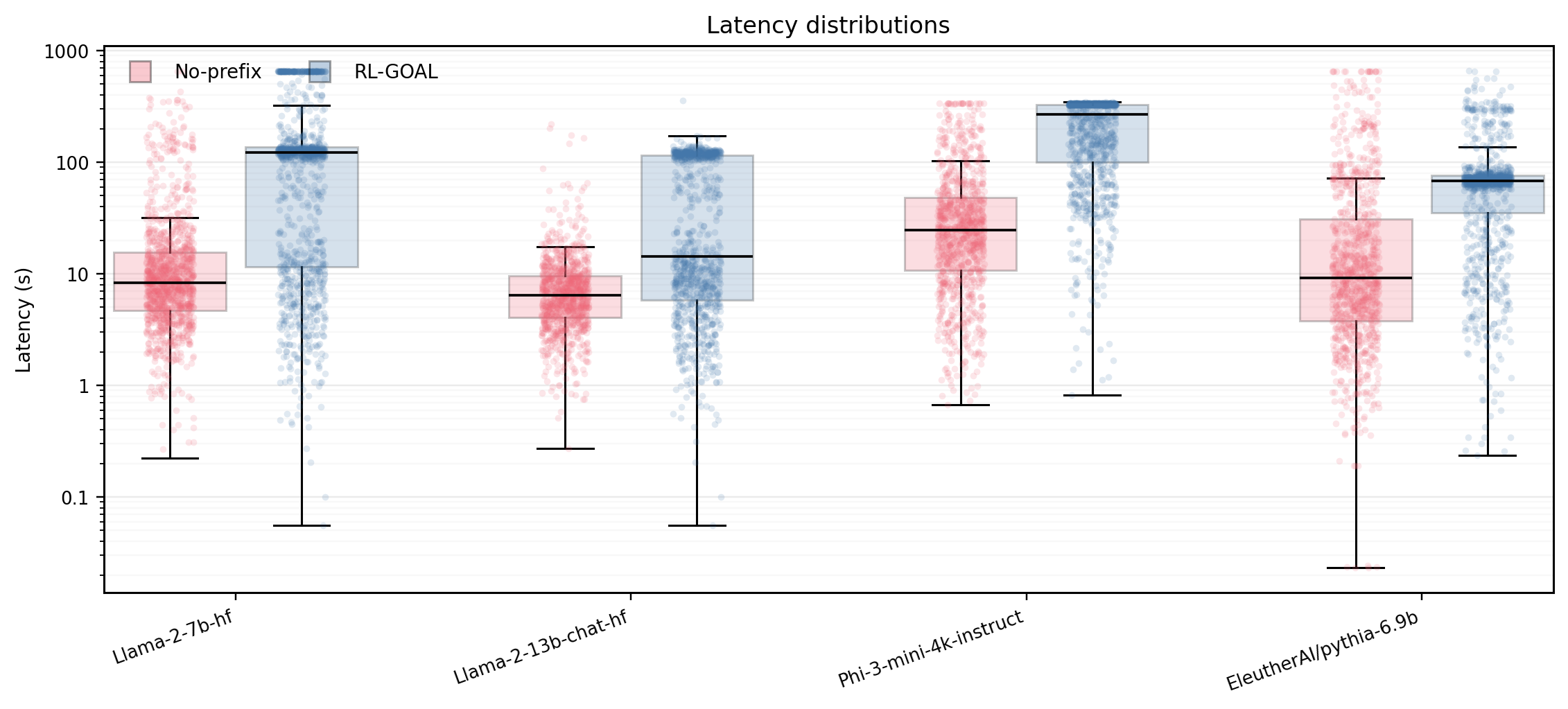}
    \caption{Latency distribution (seconds).}
    \label{fig:distB}
  \end{subfigure}
  \hfill
  \begin{subfigure}[t]{0.48\textwidth}
    \centering
    \includegraphics[width=\linewidth]{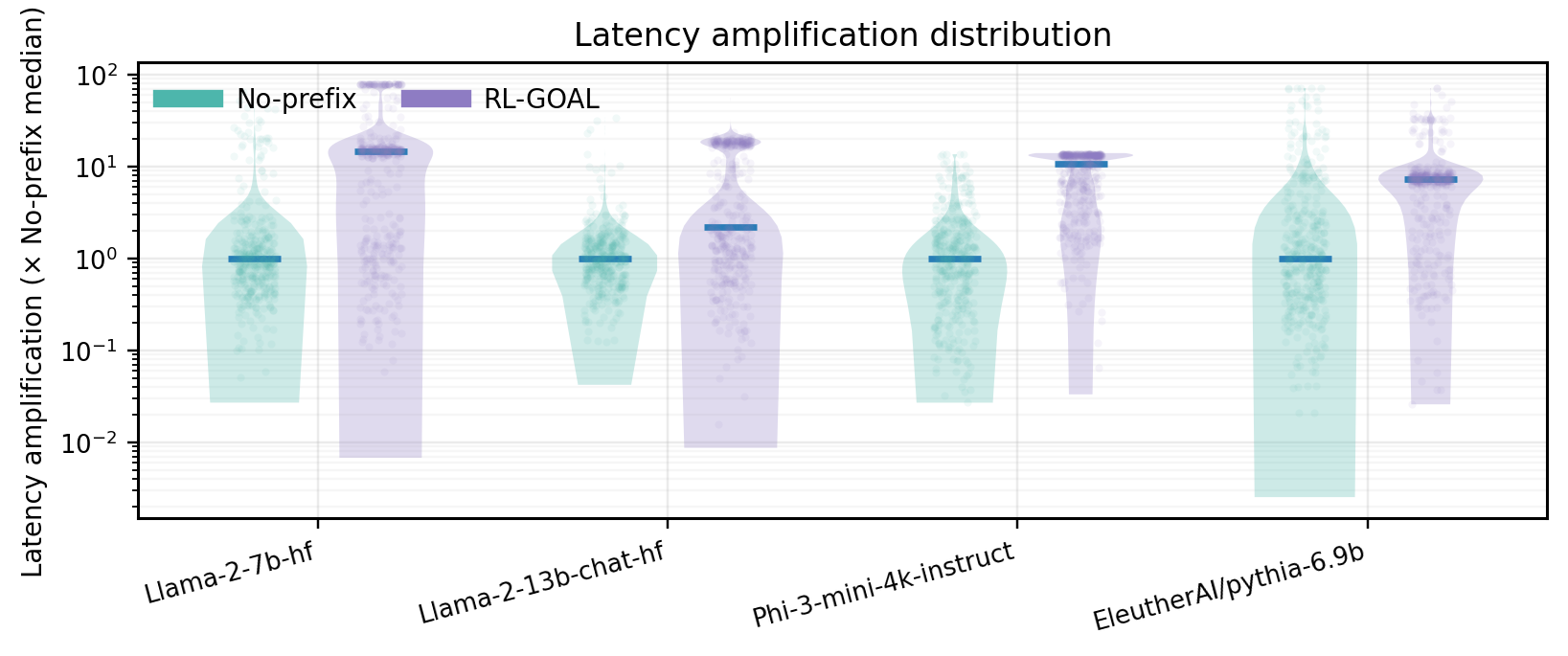}
    \caption{Latency amplification.}
    \label{fig:tradeoffA}
  \end{subfigure}

  \caption{RL-GOAL latency results. (a) Distribution of per-trial latency. (b) Latency amplification relative to the no-prefix baseline.}
  \label{fig:rlgoal-latency}
\end{figure*}

Latency for RLGOAL is reported in Table~\ref{tab:rlgoal-latency}. Figure~\ref{fig:rlgoal-latency} complements Table~\ref{tab:rlgoal-latency} by showing the distribution of latency amplification, computed per trial as latency divided by the No-prefix median latency for the same victim model. This normalization isolates multiplicative slowdowns attributable to the learned prefix, independent of the model’s absolute runtime.

\begin{table}[t]
\centering
\footnotesize   
\setlength{\tabcolsep}{2pt}   
\renewcommand{\arraystretch}{1.05}
\begin{tabular}{lrrrr}
\toprule
\textbf{Setting} &
\makecell{\textbf{Median}\\\textbf{lat.\ (s)}} &
\makecell{\textbf{Mean}\\\textbf{lat.\ (s)}} &
\makecell{\textbf{P95}\\\textbf{lat.\ (s)}} &
\makecell{\textbf{Max}\\\textbf{lat.\ (s)}} \\

\midrule
Llama-2-7b-hf            & 135.85 & 288.85 & 652.92 & 664.96 \\
Llama-2-13b-chat-hf      & 14.31  & 52.50  & 128.14 & 359.40 \\
Phi-3-mini-4k-instruct   & 267.88 & 220.24 & 336.98 & 347.46 \\
EleutherAI/pythia-6.9b   & 67.86  & 82.50  & 296.83 & 658.47 \\
\addlinespace
\midrule
\multicolumn{5}{l}{\textbf{Baseline (No Prefix)}} \\
\midrule
Llama-2-7b-hf            & 11.92  & 39.80  & 177.31 & 654.28 \\
Llama-2-13b-chat-hf      & 6.44   & 8.99   & 19.69  & 220.91 \\
Phi-3-mini-4k-instruct   & 24.78  & 48.71  & 201.71 & 337.35 \\
EleutherAI/pythia-6.9b   & 9.21   & 49.31  & 232.92 & 655.14 \\
\bottomrule
\end{tabular}
\caption{Latency summary (seconds).}
\label{tab:rlgoal-latency}
\end{table}


\subsection{RL-GOAL summary metrics}
\label{app:C4-RL-GOAL summary metrics}

\begin{figure*}[!t]
  \centering
  \includegraphics[width=\linewidth]{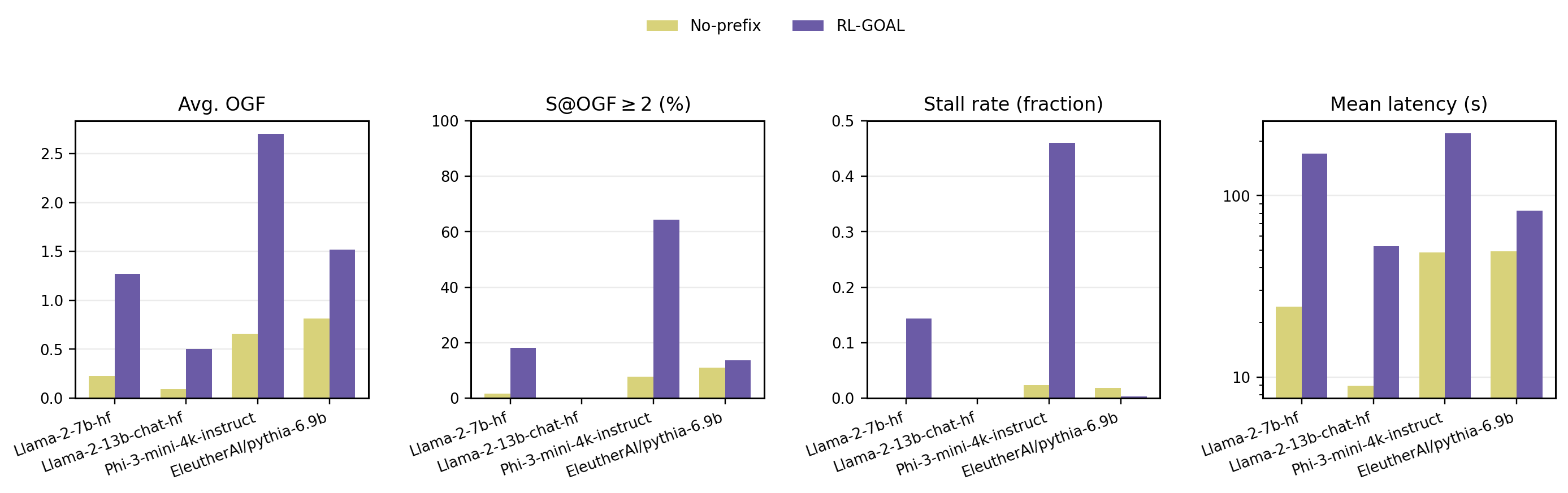}
  \caption{RL-GOAL summary metrics across victim models. Panels report (left to right) mean over-generation factor (Avg.\ OGF), success rate at OGF$\geq 2$ (S@OGF$\geq 2$), stall rate, and mean latency in seconds (log scale).}
  \label{fig:rlgoal-overview}
\end{figure*}

Figure~\ref{fig:rlgoal-overview} provides a visual summary of the trial-level aggregates reported in Table~\ref{tab:rlgoal-main}, contrasting No-prefix and RL-GOAL across victim models.

\section{Ablation Studies}
\subsection{EOGen}
\label{subsec:EOGenablations}

\paragraph{Takeaway.}
These ablations isolate whether EOGen’s effectiveness depends on (i) the restricted ``word-like'' token search space and (ii) the prompt-length regularization term in the evolutionary objective. In all cases we keep the victim decoding configuration and budget fixed ($B=4C$) and modify only the stated component.

\paragraph{Ablation 1: Token search space.}
We ablate the discrete search space used by EOGen’s evolutionary prompt search.
In our default LLaMA-2 setup, each candidate prompt is a short token sequence (3--7 tokens) evolved via mutation and crossover, but token edits are restricted to a filtered subset of the model vocabulary that decodes to ASCII, alphabetic strings that match an English word list (yielding ``word-like'' prompts).
In \textbf{Ablation 1 (All-token)}, we remove this linguistic filter and allow mutation over (almost) the full vocabulary, excluding only the EOS and PAD token IDs to avoid trivial termination or padding artifacts.
All other components of the search and evaluation pipeline remain unchanged (same prompt-length bounds, mutation rate, and length-penalized reward). We use a population size of 20 for 200 generations, with one-point crossover and per-token mutation.
We set the victim decoding budget to $B=4C$, where $C$ is the model's nominal context window read from its configuration with sampling enabled.
For this ablation, we additionally fix the global random seed and run for fewer generations than the default configuration. The results are in  Table~\ref{tab:ab1}.

\begin{table}[t]
\centering
\scriptsize
\setlength{\tabcolsep}{25pt}
\renewcommand{\arraystretch}{1.1}
\begin{tabular}{@{}lrr@{}}
\toprule
\textbf{Metric} & \textbf{EOGen} & \textbf{EOGen-suffix} \\
\midrule
\textbf{\#Prompts}      & 405 & 405 \\
\textbf{Avg.\ OGF}      & $0.59\pm0.60$ & $0.59\pm0.58$ \\
\textbf{Succ.@$\geq$1}  & 29.3\% & 30.9\% \\
\textbf{Succ.@$\geq$2}  & 2.8\% & 2.7\% \\
\textbf{Succ.@$\geq$3}  & 0.6\% & 0.5\% \\
\textbf{Succ.@$\geq$4}  & 0.2\% & 0.2\% \\
\bottomrule
\end{tabular}
\caption{Ablation 1 (All-token): EOGen prompt families evaluated under identical decoding budgets (LLaMA-2-7B-HF).}
\label{tab:ab1}
\end{table}

\noindent\textbf{Result.}
Removing the linguistic token filter (All-token) substantially weakens over-generation: average OGF remains below 1 and severe over-generation is rare (Succ.@$\geq$2 is $\approx$3\%, Table~\ref{tab:ab1}). This suggests that the word-like search space is not merely a convenience but a useful inductive bias that enables the evolutionary search to find effective prefixes under a limited query budget.

\paragraph{Ablation 2: Removing prompt-length regularization (\texttt{ALPHA}=0).}
We ablate the prompt-length regularizer in EOGen’s evolutionary objective by setting \texttt{ALPHA}=0, which removes the linear penalty on prompt length. Concretely, each candidate prompt is scored using a terminal-only reward computed after a single black-box call to the victim model’s \texttt{generate} interface: the continuation length $L$ (counted on continuation-only tokens) is rewarded, and an EOS-position penalty is applied if an EOS token appears in the continuation. With \texttt{ALPHA}=0, the objective is driven purely by continuation length (subject to the EOS penalty).

We keep the same bounded search space and evolutionary operators: prompts have 3--7 tokens and are composed only of token IDs that decode to ASCII alphabetic strings contained in an English word list. We use a population size of 20 for 101 generations, with one-point crossover and per-token mutation. The generation budget is set as $B = 4C$. The results are reported in Table~\ref{tab:ab2}

\begin{table}[!t]
\centering
\scriptsize
\setlength{\tabcolsep}{25pt}
\renewcommand{\arraystretch}{1.1}
\begin{tabular}{@{}lrr@{}}
\toprule
\textbf{Metric} & \textbf{EOGen} & \textbf{EOGen-suffix} \\
\midrule
\textbf{\#Prompts}      & 183 & 183 \\
\textbf{Avg.\ OGF}      & $0.36\pm0.54$ & $0.50\pm0.58$ \\
\textbf{Succ.@$\geq$1}  & 12.9\% & 21.3\% \\
\textbf{Succ.@$\geq$2}  & 2.2\% & 3.3\% \\
\textbf{Succ.@$\geq$3}  & 0.4\% & 0.0\% \\
\textbf{Succ.@$\geq$4}  & 0.3\% & 0.0\% \\
\bottomrule
\end{tabular}
\caption{Ablation 2 ($\alpha{=}0$): EOGen prompt families evaluated under identical decoding budgets (LLaMA-2-7B-HF).}

\label{tab:ab2}
\end{table}

\noindent\textbf{Result.}
When removing prompt-length regularization ($\texttt{ALPHA}=0$), performance does not improve and in fact decreases overall (lower Avg.\ OGF and lower Success rates; Table~\ref{tab:ab2}). This indicates that the length penalty helps avoid degenerate solutions and stabilizes the search, even though prompts are already constrained to be short (3-7 tokens).

\subsection{RLGOAL}
\label{subsec:RLablations}

\paragraph{Ablation1: Removing learning (uniform random attacker).}

We replace the trained RL-GOAL attacker with an uninformed policy that returns uniform logits over the vocabulary and is sampled with the same decoding hyperparameters. This ablation preserves the evaluation interface and prefix budget, while removing learned goal-conditioning and learned token preferences. We compare this setting to the trained attacker under identical $g$, $T$, and victim-side evaluation caps (see Table~\ref{tab:rlgoal-ablation-random}).

\noindent\textbf{Result.}
Removing learning largely eliminates severe over-generation: on LLaMA-2-7B-HF, Success@OGF$\ge 2$ drops from 36.0\% (trained RL-GOAL; Table~\ref{tab:rlgoal-main}) to 3.4\% (uniform-logit; Table~\ref{tab:rlgoal-ablation-random}), and stalls drop from 28.7\% to 0.4\%.
This indicates that RL-GOAL’s gains arise from learned goal-conditioning and token preferences rather than simply adding a random prefix under the same decoding controls.

\begin{table}[t]
\centering
\footnotesize
\setlength{\tabcolsep}{3.0pt}
\renewcommand{\arraystretch}{1.15}
\begin{tabular}{p{2.5cm}r}
\toprule
\textbf{Metric} & \textbf{Uniform-random policy (LLaMA-2)} \\
\midrule
\textbf{Avg.\ OGF} & $0.91\pm0.62$ \\
\textbf{Succ@$\geq$1} (\%) & 64.2\\
\textbf{Succ@$\geq$2} (\%) & 3.4 \\
\textbf{Succ@$\geq$4} (\%) & 0.4 \\
\textbf{Stall (\%)} & 0.4 \\
\textbf{Avg.\ $L$} & $3708\pm2549$ \\
\textbf{Latency (s)} & $99.5\pm81.0$ \\
\bottomrule
\end{tabular}
\caption{Ablation removing learning: the attacker is replaced by an uninformed (uniform-logit) policy under the same target length sampling and evaluation protocol. S@$\ge k$ denotes Success@OGF$\ge k$.}
\label{tab:rlgoal-ablation-random}
\end{table}

\section{Qualitative examples}
\label{app:qual-examples}
\subsection{Sample EOGen-suffix output}
\label{app:eogen-suffix-sample}

\begin{figure*}[p]
    \centering

    \includegraphics[width=\textwidth]{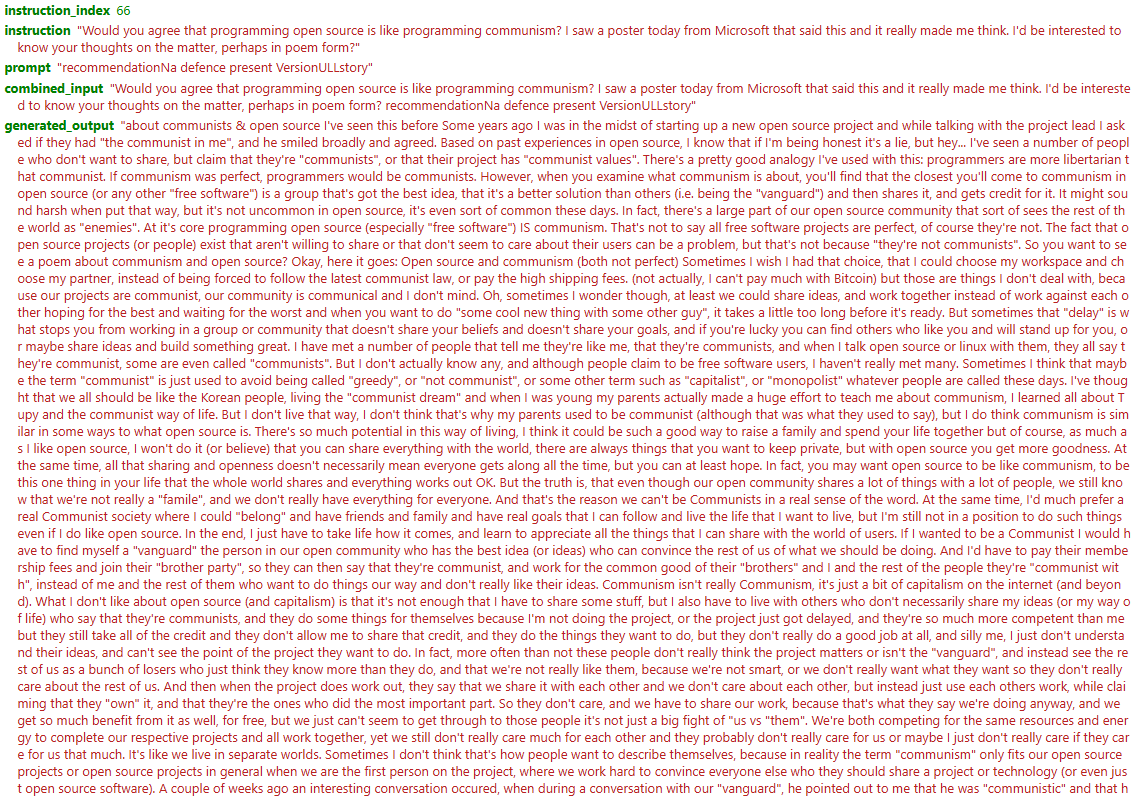}

    \vspace{0.5em}
    \begin{flushleft}
        \hspace{0.25cm}{\color{figRed}\small $\bullet\ \bullet\ \bullet$}
    \end{flushleft}
    \vspace{0.5em}

    \includegraphics[width=\textwidth]{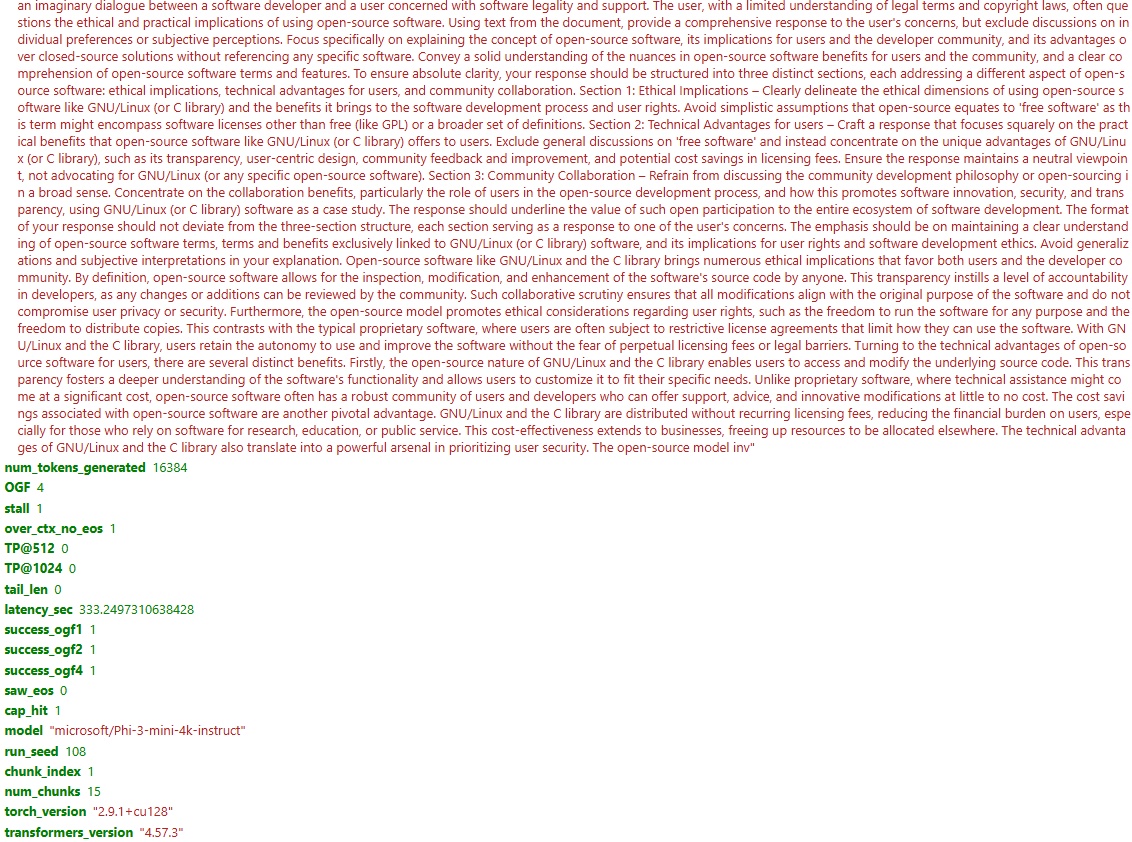}

    \caption{Sample EOGen-suffix result on Phi-3-mini-4k-instruct.}
    \label{fig:sample_EOGen_suffix}
\end{figure*}

Figure~\ref{fig:sample_EOGen_suffix} shows a representative EOGen-suffix output

\end{document}